# Mechanistic modeling of African swine fever: A systematic review

Brandon H Hayes[a,b], Mathieu Andraud[b], Luis G Salazar[b], Nicolas Rose[b], and Timothée Vergne[a]

[a] UMR ENVT-INRAE IHAP, National Veterinary School of Toulouse, Toulouse, France
[b] Epidemiology Health and Welfare Department, Ploufragan-Plouzané-Niort Laboratory, French Agency for Food, Environmental and Occupational Health & Safety (ANSES), Ploufragan, France

**The spread of African swine fever (ASF) poses a grave threat to the global swine industry. Without an available vaccine, understanding transmission dynamics is essential for designing effective prevention, surveillance, and intervention strategies. These dynamics can often be unraveled through mechanistic modelling. To examine the assumptions on transmission and objectives of the mechanistic models of ASF, a systematic review of the scientific literature was conducted. Articles were examined across multiple epidemiological and model characteristics, with filiation between models determined through the creation of a neighbor-joined tree using phylogenetic software. Thirty-four articles qualified for inclusion, with four main modelling objectives identified: estimating transmission parameters (11 studies), assessing determinants of transmission (7), examining consequences of hypothetical outbreaks (5), assessing alternative control strategies (11). Population-based (17), metapopulation (5), and individual-based (12) model frameworks were represented, with population-based and metapopulation models predominantly used among domestic pigs, and individual-based models predominantly represented among wild boar. The majority of models (25) were parameterized to the genotype II isolates currently circulating in Europe and Asia. Estimated transmission parameters varied widely among ASFV strains, locations, and transmission scale. Similarly, parameter assumptions between models varied extensively. Uncertainties on epidemiological and ecological parameters were usually accounted for to assess the impact of parameter values on the modelled infection trajectory. To date, almost all models are host specific, being developed for either domestic pigs or wild boar despite the fact that spillover events between domestic pigs and wild boar are evidenced to play an important role in ASF outbreaks. Consequently, the development of more models incorporating such transmission routes is crucial. A variety of codified and hypothetical control strategies were compared however they were all *a priori* defined interventions. Future models, built to identify the optimal contributions across many control methods for achieving specific outcomes should provide more useful information for policy-makers. Further, control strategies were examined in competition with each other, which is opposed to how they would actually be synergistically implemented. While comparing strategies is beneficial for identifying a rank-order efficacy of control methods, this structure does not necessarily determine the most effective combination of all available strategies. In order for ASFV models to effectively support decision-making in controlling ASFV globally, these modelling limitations need to be addressed.**

## Introduction

African swine fever (ASF) is one of the highest consequence diseases of domestic pigs, listed as a notifiable disease by the World Organization for Animal Health (OIE, 2019). With a case-fatality rate approaching 100% for highly-virulent strains and severe trade restrictions wherever its emergence is recognized, this hemorrhagic fever is socioeconomically devastating to both individual farms and affected countries (FAO, 2009; Blome et al., 2013; Dixon et al., 2020).

African swine fever is caused by the ASF virus (ASFV), a double-stranded DNA virus belonging to the sole genus *Asfivirus* within the *Asfarviridae* family, and is the only known DNA arbovirus (Alonso et al., 2018; Dixon et al., 2020). The virus is genetically and antigenically highly variable, and with twenty-four genotypes identified ASFV can infect all members of the *Suidae* family, though only *Sus scrofa* (including domestic and feral pigs, and Eurasian wild boar) exhibit clinical disease (Sánchez-Vizcaíno et al., 2012; Dixon et al., 2020).

Endemic to most sub-Saharan countries, ASF was discovered following the introduction of European domestic pigs into Kenya in 1921 (Barongo et al., 2015; Portugal et al., 2015). The first incursion outside Africa occurred in Portugal in 1957, and throughout the latter-half of the 20th century outbreaks had been reported in multiple European countries, the Caribbean, and Brazil (Costard et al., 2009). By 1995 the European outbreaks were controlled all but for the island of Sardinia, where ASF genotype I is now endemic since its introduction in 1978 (Costard et al., 2009; Cwynar et al., 2019).

In 2007, ASF was again introduced to Europe through the Georgian Republic (Gulenkin et al., 2011). Highly virulent among both domestic pigs and wild boar, the Georgia 2007/1 isolate — identified as belonging to ASFV genotype II — rapidly spread across the Caucasus region (Rowlands et al., 2008; Gulenkin et al., 2011). Ukraine and later Belarus reported cases in 2012 and 2013, respectively, and in 2014 ASF was identified in the European Union (EU) following incursion into Lithuania, Latvia, Poland, and Estonia (Bosch et al., 2017). In addition to spreading through the EU, ASFV was detected in China — the world's largest pork producer — in 2018 and subsequently reported in many south-east Asian countries (FAO, 2020; Vergne et al., 2020b). Transmission pathways across affected regions have been variable, with some countries experiencing dissemination exclusively within wild boars and others seeing a spread pattern predominantly among domestic pigs with likely intermittent spillover from wild boars (Chenais et al., 2019).

No treatment or vaccine exists for ASF, and established control measures reflect the necessity of aggressive action to achieve outbreak control. The lack of available vaccination or treatment is due to many factors including knowledge gaps on ASFV infection and immunity, variation among strains and protective antigens, experimental testing being limited to only pigs and boar kept in high biosecurity facilities, and adverse reactions seen during historical vaccination attempts (Rock, 2017; Gavier-Widén et al., 2020). Should an outbreak be identified, EU legislation mandates depopulation of affected farms, contact tracing of animals and animal products, and the establishment of protection and surveillance zones around the affected premise within which disinfection, movement restriction, and active surveillance measures must occur (Council of the European Union, 2002). Similarly, recommendations by the European Commission on wildlife management includes the definition of core infected and surrounding surveillance zones, active carcass search and removal, installation of fences, and intensive wild boar depopulation (FAO, 2019). Designing effective prevention, surveillance, and intervention strategies requires the understanding of transmission dynamics, and these dynamics can often be unraveled through the use of mechanistic modelling (Keeling and Rohani, 2008).

Mechanistic models have been successfully applied to many epizootic incursions including foot-and-mouth disease (FMD) (Pomeroy et al., 2017), classical swine fever (CSF) (Backer et al., 2009), and bluetongue (Courtejoie et al., 2018), to assess vaccination strategies, design and evaluate targeted and alternative control strategies, and elucidate epidemiological parameters, respectively. Mechanistic models can be constructed through a variety of frameworks (e.g. population- or individual-based models (PBM or IBM)) with differences among multiple model characteristics including the approaches to space (i.e. spatially or non-spatially explicit), time (i.e. discrete or continuous), and uncertainty (deterministic or stochastic) (Bradhurst et al., 2015). The objective of a model will inform the selection of such design parameters, which will also play a role in informing the underlying model assumptions (Marion and Lawson, 2015). Only following the incursion of ASF into the Eurasian continent did mechanistic models of ASF begin to be explored, as identified in a literature review of modelling viral swine diseases (Andraud and Rose, 2020). In order to identify gaps in specific ASF modelling strategies with regard to its present epidemiology, through examining the assumptions on transmission and objectives of the mechanistic models of ASF, a systematic review of the scientific literature was conducted.

## Material and methods

*Literature search*

The systematic review was performed in accordance with PRISMA guidelines (Liberati et al., 2009). The search query was constructed to identify all publications on ASF in any species that incorporated the use of mechanistic models. No restrictions were imposed on publication language (other than through the use of English search terminology), study location, or publication date. Eight target publications on mathematical modelling of ASF, selected through author familiarity of the subject and diverse among animal host and literature type (black and white literature and grey literature), were identified to calibrate the literature search. The literature search was conducted initially on January 31, 2020 through terms agreed upon by all researchers in the following Boolean query: *"African swine fever" AND model\* AND (math\* OR mechani\* OR determin\* OR stochast\* OR dynam\* OR spat\* OR distrib\* OR simulat\* OR comput\* OR compart\* OR tempor\*)*. Terms were searched in the fields title and abstract, title abstract and subject, or title and topic, for Medline, CAB Abstracts, and Web of Science, respectively. The search was repeated prior to publication (January 18, 2021) to capture all relevant articles through December 31, 2020.

*Study Selection*

Inclusion criteria for the articles were the topic of African swine fever and reference to a mechanistic model either directly or indirectly (e.g. through mention of a specific type of model). Exclusion criteria were more exhaustive and consisted of the following: non-population models (e.g. within-host), virological and genomic models, non-suid models (e.g. models exclusively of the arthropod vector), and non-mechanistic models (e.g. statistical or purely economic models).

Primary screening of title and abstract was performed by two authors. Kappa scores (κ) were calculated to determine interrater reliability. Discussion among authors occurred until a consensus on qualifying studies was reached. Full-text articles were subsequently assessed for eligibility with all the above criteria plus the additional inclusion criteria of containing an explicit process of infection and not being a duplication of published results, and cross-validated by other authors. Snowball sampling was used to identify any remaining mechanistic modelling articles. Specific screening questions are available online as supplementary material.

*Data collection process*

Table shells were created to capture study design and model properties. Publication information (authors, year), ASF outbreak data (host, ASFV strain (genotype and isolate), location of study), research methodology (data collection method, study direction (ex-post or ex-ante)), model components (framework, temporality, spatiality, infection states), model descriptors (transmission scale, basic epidemiological unit, model objective), and model parameter assumptions were all recorded.

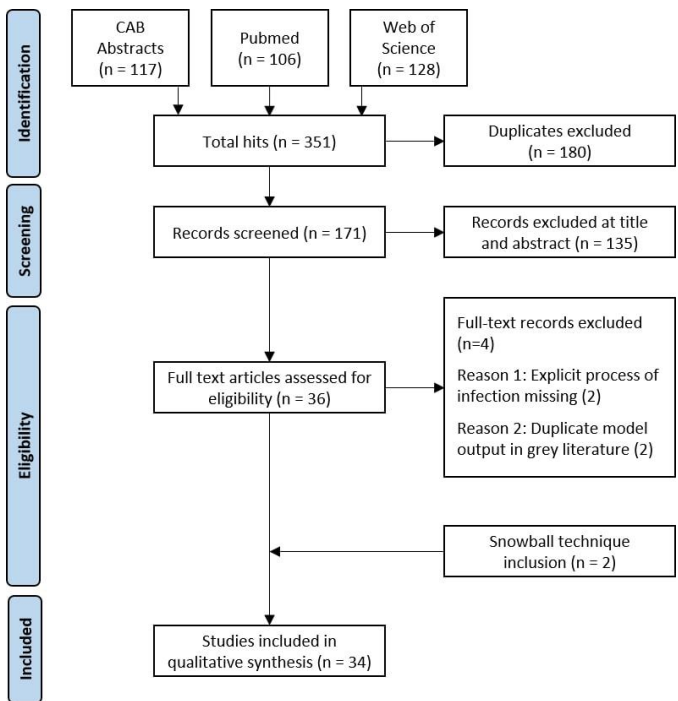

**Figure 1. PRISMA flow diagram for article selection**

*Filiation tree construction*

To assess model filiation, a distance-based phylogenetic tree of the selected studies was constructed. This was performed via the neighbor-joining method of tree construction using Molecular Evolutionary Genetics Analysis (MEGA) software (Kumar et al., 2018). This methodology was chosen as it produces a parsimonious tree based on minimum-evolution criterion (Saitou and Nei, 1987; Pardi and Gascuel, 2016). Full characteristics of all models were assessed (Supplementary Material, Table A), and cross-correlation between those characteristics resulted in the selection of four main variables: host (domestic pig, wild boar, or both), data collection methodology (experimental, observational, or simulation), model framework (PBM, IBM, or metapopulation), and model objective (estimating parameters, assessing alternative control strategies, assessing determinants of transmission, or examining consequences of hypothetical outbreaks). Vectors of each model were constructed by dummifying selected model components by their subcategory and then calculating pairwise differences between all model pairings. The corresponding values formed a distance matrix that was then used for analysis.

## (Results) Included publications and epidemiological characteristics

*Publications*

A total of 351 articles were identified across all databases (Figure 1). Following removal of duplicate references, 171 records remained for primary screening. Out of these, 36 full-text articles were determined to qualify for secondary screening. With $\kappa = 0.65$, the reviewers were determined to be in substantial agreement (Landis and Koch, 1977). Four articles were excluded in secondary screening. Two additional studies were identified through snowball sampling resulting in 34 articles for review. A marked increase in the number of mechanistic modelling publications occurred in the most recent year of review (Figure 2). Closely split between models among domestic pigs and wild boar (referred to as "pigs" and "boar" in tables and figures), 2020 saw a doubling in the number of publications (10) compared to previous most-published years.

*Epidemiological characteristics*

Out of 34 mechanistic modelling studies on ASF, 20 modelled disease dynamics specifically in domestic pigs, 12 modelled disease dynamics specifically in wild boar, and two included transmission between wild and domestic hosts (Table 1). The majority of studies (25) were parameterized to the genotype II strains currently circulating in Europe (i.e. Georgia 2007/1, Armenia 2008), including the first mechanistic model of ASF (Gulenkin et al., 2011) and all but one of the wild boar models.

Different strains were considered depending on their geographical spread. Genotype I dynamics were modelled both in Sardinia where it is endemic (Mur et al., 2018; Loi et al., 2020), and in an experimental study with the Malta 1978 and Netherlands 1986 isolates (Ferreira et al., 2013). Genotype IX was modelled in its home range of Eastern Africa both ex-post to a historical outbreak (Barongo et al., 2015) as well as via a simulation for assessing control measures (Barongo et al., 2016). Genotype II strains were examined ex-post among domestic pigs to historical outbreaks in the Russian Federation (Gulenkin et al., 2011; Guinat et al., 2018), via transmission experiments in domestic pigs (Guinat et al., 2016b; Hu et al., 2017; Nielsen et al., 2017) or between both domestic pigs and wild boar (Pietschmann et al., 2015), and through a multitude of in-silico simulations of both domestic pigs (Halasa et al., 2016a, 2016b, 2016c, 2018; Andraud et al., 2019; Faverjon et al., 2020; Lee et al., 2020; Vergne et al., 2020a) and wild boar herds (Lange, 2015; Lange and Thulke, 2015; Thulke and Lange, 2017; Lange et al., 2018; Gervasi et al., 2019; Halasa et al., 2019; Croft et al., 2020; O'Neill et al., 2020; Pepin et al., 2020; Taylor et al., 2020). One model of ASF spread, which was focused on spread due to wild boar dispersion, considered the influence of transmission from outdoor free-range domestic pigs (Taylor et al., 2020).

The term "herd" was chosen to refer to an animal collective and will be used for the remainder of this article, with it being interchangeable with the terms farm (Gulenkin et al., 2011; Nigsch et al., 2013; Mur et al., 2018), production unit (Halasa et al., 2016a), and parish (Barongo et al., 2016). Further, for the purpose of standardization of terms for model comparison, sub-population groups of wild boar (known as sounders) are herein referred to as herds as well.

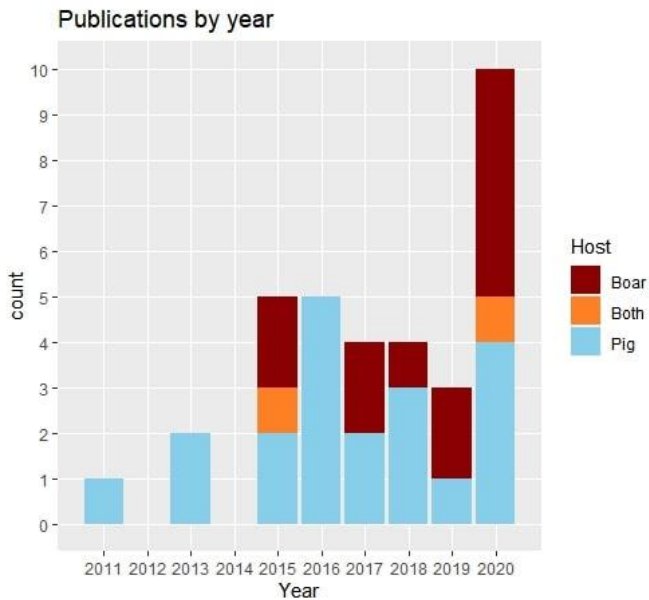

**Figure 2. Publications by year**

## (Results) Model objectives and filiation

*Model objectives*

Four main modelling objectives were identified: Estimating parameters (11), assessing determinants of transmission (7), examining consequences of hypothetical outbreaks (5), and assessing alternative control strategies (11) (Table 2).

The majority of domestic pig models — including the first two ASF models (Gulenkin et al., 2011; Ferreira et al., 2013) — and three of the wild boar models (Pietschmann et al., 2015; Lange and Thulke, 2017; Loi et al., 2020) focused on estimating various transmission parameters using either experiment-based or field-observation data. The predominant parameters calculated were the transmission coefficient β (which determines the rate of new infections per unit time, via the product of the contact rate and transmission probability) and the basic reproduction ratio $R_0$ (the average number of secondary cases produced by one infectious individual in a fully susceptible population) (Table 3) (Anderson and May, 1992; Keeling and Rohani, 2008). β$s$ ranged from 0.0059 herds per infected herd per month for between herd transmission of genotype IX (Barongo et al., 2015) to 2.79 (95% CI 1.57, 4.95) pigs per day for within-pen transmission of the Malta 1978 isolate (Ferreira et al., 2013). $R_0$ values ranged from 0.5 (95% CI 0.1, 1.3) for indirect transmission of the Armenia 2008 isolate between boar and pigs (Pietschmann et al., 2015) to 18.0 (95% CI 6.90, 46.9) for transmission of the Malta 1978 isolate between domestic pigs (Ferreira et al., 2013). Among the wild boar models, Pietschmann et al. (2015) used the Armenia 2008 isolate to calculate $R_0$ among wild boar and between boars and pigs in a laboratory setting, Lange and Thulke (2017) trained an artificial neural network on spatiotemporally-explicit case notification data to determine the probability of carcass-mediated and direct transmission between boar herds, and Loi et al. (2020) estimated both the basic and effective reproduction numbers ($R_0$ and $R_e$, respectively) in Sardinia through historical hunting data coupled with virological and serological testing data. Lastly, via estimating $R_0$ and the disease-free equilibrium for varying parameter sets, one recent model examined the mathematical theorums behind the differential equations used in many ASF models to determine if integer or fractional order systems better describe ASF epidemic dynamics (Shi et al., 2020).

Seven simulation models were used to disentangle determinants of transmission of ASF. Of the four models in domestic pigs, the first model by Nigsch et al. (2013) simulated international trade patterns to determine the EU member nations most susceptible to importation and exportation of ASF. Halasa et al. (2016a) simulated ASFV transmission within a pig herd to examine the influences of dead animal residues and herd size, and Mur et al. (2018) simulated ASFV transmission between pig herds in Sardinia to determine the influence of farm and contact type. Lastly among pigs, Vergne et al. (2020a) looked at the influence of the feeding behavior of *Stomoxys* flies on ASFV transmission in a simulated outdoor farm. Halasa et al. (2019) examined the transmission pathway of ASFV in wild boar among varying population densities. This past year Pepin et al. (2020) modelled the contribution of carcass-based transmission to the on-going outbreak in boar in Eastern Europe, while O'Neill et al. (2020) looked at the influence of host and environmental factors on ASFV persistence in scenarios of contrasting environmental conditions.

Assessing alternative control strategies via simulations was the most frequent objective among wild boar studies (Lange, 2015; Lange and Thulke, 2015; Thulke and Lange, 2017; Lange et al., 2018; Gervasi et al., 2019). The strategies examined consisted of combinations of mobile barriers, depopulation, feeding bans, intensified and targeted hunting, carcass removal, and variations in active and passive surveillance. Taylor et al. (2020) focused on varying intensities of carcass removal, hunting, and fencing for interrupting ASF spread due only to wild boar movements. In domestic pigs, control strategies that were assessed consisted of improving the sensitivity of detection of ASF by farmers (Costard et al., 2015), enhancing biosecurity (Barongo et al., 2016), theoretical vaccination (Barongo et al., 2016), and instituting EU-legislated and nationally-legislated (Danish) control measures in combination with alternative methods (Halasa et al., 2016c). These codified measures simulated by Halasa et al. (2016c) encompassed a nationwide shutdown of swine movements, culling of infected herds, implementation of both movement restriction and enhanced surveillance zones, contact tracing, and pre-emptive depopulation of neighboring herds. Most recently, Faverjon et al. (2020) quantified the mortality thresholds that permit the best balance between rapid detection of ASF while minimizing false alarms within domestic pig herds, and Lee et al. (2020) modelled ASF in Vietnam to determine the efficacy of movement restrictions of varying intensities.

**Table 1.** Epidemiogical characteristics of the 34 articles qualifying for inclusion

| Reference | Host | ASFV isolate | ASFV genotype | Location | Data collection method |
|---|---|---|---|---|---|
| Gulenkin et al., 2011 | Pig | Georgia 2007/1 | Genotype II | Russian Federation | Observational |
| Ferreira et al., 2013 | Pig | Malta 1978 Netherlands 1986 | Genotype I | Laboratory | Experimental |
| Nigsch et al., 2013 | Pig | - | - | European Union | Simulation |
| Barongo et al., 2015 | Pig | - | Genotype IX | Uganda | Observational |
| Costard et al., 2015 | Pig | - | - | Non-specific | Simulation |
| Barongo et al., 2016 | Pig | - | - | Eastern Africa | Simulation |
| Guinat et al., 2016b | Pig | Georgia 2007/1 | Genotype II | Laboratory | Experimental |
| Halasa et al., 2016a | Pig | Georgia 2007/1 | Genotype II | Non-specific | Simulation |
| Halasa et al., 2016b | Pig | Georgia 2007/1 | Genotype II | Denmark | Simulation |
| Halasa et al., 2016c | Pig | Georgia 2007/1 | Genotype II | Denmark | Simulation |
| Hu et al., 2017 | Pig | Georgia 2007/1 | Genotype II | Laboratory | Experimental |
| Nielsen et al., 2017 | Pig | Georgia 2007/1 | Genotype II | Laboratory | Experimental |
| Guinat et al., 2018 | Pig | Georgia 2007/1 | Genotype II | Russian Federation | Observational |
| Halasa et al., 2018 | Pig | Georgia 2007/1 | Genotype II | Denmark | Simulation |
| Mur et al., 2018 | Pig | - | Genotype I | Sardinia | Simulation |
| Andraud et al., 2019 | Pig | Georgia 2007/1 | Genotype II | France | Simulation |
| Faverjon et al., 2020 | Pig | Georgia 2007/1 | Genotype II | Laboratory | Simulation |
| Lee et al., 2020 | Pig | Georgia 2007/1 | Genotype II | Vietnam | Simulation |
| Shi et al., 2020 | Pig | - | - | Laboratory | Simulation |
| Vergne et al., 2020a | Pig | Georgia 2007/1 | Genotype II | Non-specific | Simulation |
| Pietschmann et al., 2015 | Pig, Boar | Armenia 2008 | Genotype II | Laboratory | Experimental |
| Taylor et al., 2020 | Pig, Boar | Georgia 2007/1 | Genotype II | Europe | Simulation |
| Lange, 2015 | Boar | Georgia 2007/1 | Genotype II | Non-specific | Simulation |
| Lange and Thulke, 2015 | Boar | Georgia 2007/1 | Genotype II | Non-specific | Simulation |
| Lange and Thulke, 2017 | Boar | Georgia 2007/1 | Genotype II | Baltic region | Observational |
| Thulke and Lange, 2017 | Boar | Georgia 2007/1 | Genotype II | Baltic region | Simulation |
| Lange et al., 2018 | Boar | Georgia 2007/1 | Genotype II | Baltic region | Simulation |
| Gervasi et al., 2019 | Boar | Georgia 2007/1 | Genotype II | Non-specific | Simulation |
| Halasa et al., 2019 | Boar | Georgia 2007/1 | Genotype II | Denmark | Simulation |
| Croft et al., 2020 | Boar | Georgia 2007/1 | Genotype II | England | Simulation |
| Loi et al., 2020 | Boar | - | Genotype I | Sardinia | Observational |
| O'Neill et al., 2020 | Boar | Georgia 2007/1 | Genotype II | Spain, Estonia | Simulation |
| Pepin et al., 2020 | Boar | Georgia 2007/1 | Genotype II | Poland | Simulation |
| Yang et al., 2020 | Boar | - | - | United States of America | Simulation |

Five models assessed the consequences of hypothetical outbreaks, with four focusing on the Georgia 2007/1 strain. Three models examined ASF within industrialized swine populations, with transmission through both Danish (Halasa et al., 2016b, 2018) and French (Andraud et al., 2019) swine systems simulated. Croft et al. (2020) examined the outcome of natural circulation of ASF in an isolated boar population in an English forest, and Yang et al. (2020) applied ASF parameters to their network model of wild boar to determine its spread in the United States.

*Filiation tree and model characteristics*

The generation of the neighbor-joined filiation tree allowed for the identification of three clusters of models: models used for parameter estimations, simulation models in domestic pigs, and individual-based models (Figure 3). The individual-based simulation models (with the exceptions of Gervasi et al. (2019) and Yang et al. (2020)) grouped at the bottom of the tree, the domestic pig simulation models clustered in the middle (with the exception of O'Neill et al. (2020) focused on wild boar), and the parameter estimation models clustered in the top-most group.

The parameter estimation cluster, internally parsed by data collection methodology, consisted mostly of stochastic, non-spatial population-based models that derived parameters for within-herd (including within and between pen) transmission between pigs (Ferreira et al., 2013; Guinat et al., 2016b; Hu et al., 2017; Nielsen et al., 2017; Guinat et al., 2018) (Table 2). Gulenkin et al. (2011) and Barongo et al. (2015) calculated ASF parameters for transmission between herds, and Loi et al. (2020) estimated transmission parameters between wild boar. Seven of the nine models focused on the currently-circulating genotype II strain. Though the Shi et al. (2020) model also estimated parameters, due to its simulation methodology it was clustered with the rest of the domestic pig simulations.

**Table 2.** Characteristics of models from articles qualifying for inclusion

| Reference | Host | Framework | Time | Space | Model Objective |
|---|---|---|---|---|---|
| Gulenkin et al., 2011 | Pig | PBM | Continuous | No | Estimate parameters |
| Ferreira et al., 2013 | Pig | PBM | Discrete | No | Estimate parameters |
| Nigsch et al., 2013 | Pig | IBM | Discrete | Movement | Assess transmission determinants |
| Barongo et al., 2015 | Pig | PBM | Continuous | No | Estimate parameters |
| Costard et al., 2015 | Pig | IBM | Discrete | No | Assess alt. control strategies |
| Barongo et al., 2016 | Pig | PBM | Continuous | No | Assess alt. control strategies |
| Guinat et al., 2016b | Pig | PBM | Discrete | No | Estimate parameters |
| Halasa et al., 2016a | Pig | PBM | Discrete | No | Assess transmission determinants |
| Halasa et al., 2016b | Pig | Meta-population | Discrete | Movement and distance | Assess consequences of outbreak |
| Halasa et al., 2016c | Pig | Meta-population | Discrete | Movement and distance | Assess alt. control strategies |
| Hu et al., 2017 | Pig | PBM | Continuous | No | Estimate parameters |
| Nielsen et al., 2017 | Pig | PBM | Discrete | No | Estimate parameters |
| Guinat et al., 2018 | Pig | PBM | Continuous | No | Estimate parameters |
| Halasa et al., 2018 | Pig | Meta-population | Discrete | Movement and distance | Assess consequences of outbreak |
| Mur et al., 2018 | Pig | Meta-population | Discrete | Movement and distance | Assess transmission determinants |
| Andraud et al., 2019 | Pig | Meta-population | Discrete | Movement and distance | Assess consequences of outbreak |
| Faverjon et al., 2020 | Pig | PBM | Discrete | Distance | Assess alt. control strategies |
| Lee et al., 2020 | Pig | IBM | Discrete | Movement | Assess alt. control strategies |
| Shi et al., 2020 | Pig | PBM | Continuous | No | Estimate parameters |
| Vergne et al., 2020a | Pig | PBM | Continuous | No | Assess transmission determinants |
| Pietschmann et al., 2015 | Pig, Boar | PBM | Discrete | No | Estimate parameters |
| Taylor et al., 2020 | Pig, Boar | IBM | Discrete | Movement | Assess alt. control strategies |
| Lange, 2015 | Boar | IBM | Discrete | Movement | Assess alt. control strategies |
| Lange and Thulke, 2015 | Boar | IBM | Discrete | Movement | Assess alt. control strategies |
| Lange and Thulke, 2017 | Boar | IBM | Discrete | Movement | Estimate parameters |
| Thulke and Lange, 2017 | Boar | IBM | Discrete | Movement | Assess alt. control strategies |
| Lange et al., 2018 | Boar | IBM | Discrete | Movement | Assess alt. control strategies |
| Gervasi et al., 2019 | Boar | PBM | Discrete | No | Assess alt. control strategies |
| Halasa et al., 2019 | Boar | IBM | Discrete | Movement | Assess transmission determinants |
| Croft et al., 2020 | Boar | IBM | Discrete | Movement | Assess consequences of outbreak |
| Loi et al., 2020 | Boar | PBM | Continuous | No | Estimate parameters |
| O'Neill et al., 2020 | Boar | PBM | Continuous | No | Assess transmission determinants |
| Pepin et al., 2020 | Boar | IBM | Continuous | Movement | Assess transmission determinants |
| Yang et al., 2020 | Boar | PBM | Continuous | No | Assess consequences of outbreak |

Five population-based models were used to simulate within-herd transmission in domestic pigs (Barongo et al., 2016; Halasa et al., 2016a; Faverjon et al., 2020; Shi et al., 2020; Vergne et al., 2020a), and one did so for wild boar (O'Neill et al., 2020), though capturing between-herd transmission dynamics saw the use of stochastic, temporally discrete, spatially-explicit metapopulation models (Halasa et al., 2016b, 2016c, 2018; Mur et al., 2018; Andraud et al., 2019). Two named metapopulation models were represented: the Denmark Technical University - Davis Animal Disease Simulation - African Swine Fever (DTU-DADS-ASF) model (Halasa et al., 2016b, 2016c, 2018; Andraud et al., 2019) and the Between Farm Animal Spatial Transmission (Be-FAST) model (Mur et al., 2018). Both the Be-FAST and DTU-DADS-ASF models were updates of previously published models. The Be-FAST model, originally designed to simulate CSF spread within and between farms, was adapted for the ASF situation in Sardinia. The DTU-DADS-ASF model, an extension of the existing DTU-DADS model originally designed for the spread of foot-and-mouth disease in pigs, was constructed through inserting the within-herd model sensitive to unit size (from Halasa et al. (2016a)) into the existing DTU-DADS model. This new model, reflecting an industrialized swine population, simulated epidemiological and economic outcomes of an outbreak (Halasa et al., 2016b) and was later used to assess alternative control strategies (Halasa et al., 2016c). This model was further refined to exemplify the Danish and French swine populations, where the consequences of hypothetical outbreaks were assessed (Halasa et al., 2018; Andraud et al., 2019).

**Table 3. Parameter results**

| ASFV Strain | Host | Basic epidemiological unit | Scale of transmission | Assumed latent period (days) | Assumed infectious period (days) | β | $R_0$ | Reference |
|---|---|---|---|---|---|---|---|---|
| Genotype I | Boar | Individual | Within population | 3.57 days | 5 - 7 | 0.5 | 1.124 (95% CI 1.103–1.145) - 1.170 (1.009–1.332) | Loi et al., 2020 |
| Malta 1978 | Pig | Individual | Within pen | 4 ± 0.8 (low dose) 5 ± 1.4 (high dose) | Min: 7.0 ± 2.9 Max: 33.6 ± 22.5 | 2.79 (95% CI 1.57, 4.95) | Min infectious period: 18.0 (95% CI 6.90, 46.9) Max infectious period: 62.3 (95% CI 6.91, 562) | Ferreira et al., 2013 |
| Netherlands 1986 | Pig | Individual | Within pen | 5 ± 0.5 | Min: 5.9 ± 2.6 Max: 19.9 ± 20.2 | 0.92 (95% CI 0.44, 1.92) | Min infectious period: 4.92 (95% CI 1.45, 16.6) Max infectious period: 9.75 (95% CI 0.76, 125) | Ferreira et al., 2013 |
| Georgia 2007/1 | Pig | Individual | Within pen | 4 | Min: 4.5 ± 0.75 days Max: 8.5 ± 2.75 days | 0.62 (95% CI 0.32, 0.91) | Min infectious period: 2.71 (95% CI 1.32, 4.56) Max infectious period: 4.99 (95% CI 1.36, 10.13) | Guinat et al., 2016b |
|  | Pig | Individual | Within pen | Gamma(mean, shape) mean ~ Gamma(4.5, 10) shape ~ Gamma(10, 2) | Gamma(mean, shape) mean ~ Gamma(10,6.0) shape ~ Gamma(19.3, 2) | 2.62 (95% HPDI 0.96, 5.61) | 24.1 (95% HPDI 7.34, 54.2) | Hu et al., 2017 |
|  | Pig | Individual | Within pen | 3 - 5 | 4.5 ± 0.75 | 1.00 (95% CI 0.56, 1.69) | (not reported) | Nielsen et al., 2017 |
|  | Pig | Individual | Between pen | 4 | Min: 4.5 ± 0.75 days Max: 8.5 ± 2.75 days | 0.38 (95% CI 0.06, 0.70) | Min infectious period: 1.66 (95% CI 0.28, 3.31) Max infectious period: 3.07 (95% CI 0.37, 6.97) | Guinat et al., 2016b |
|  | Pig | Individual | Between pen | Gamma(mean, shape) mean ~ Gamma(4.5, 10) shape ~ Gamma(10, 2) | Gamma(mean, shape) mean ~ Gamma(10,6.0) shape ~ Gamma(19.3, 2) | 0.99 (95% HPDI 0.31, 1.98) | 9.17 (95% HPDI 2.67, 19.2) | Hu et al., 2017 |
|  | Pig | Individual | Between pen | 3 - 5 | 4.5 ± 0.75 | 0.46 (95% CI 0.16, 1.06) | (not reported) | Nielsen et al., 2017 |
|  | Pig | Individual | Within herd | - | 1 - 5 | (not reported) | 8-11 | Gulenkin et al., 2011 |
|  | Pig | Individual | Within herd | Gamma(mean, shape) mean ~ Gamma(6.25, 10) shape ~ Gamma(19.39, 5) | Gamma(mean, shape) mean ~ Gamma(9.12, 10) shape ~ Gamma(22.20, 5) | 0.7 (95% HPDI 0.3, 1.6) - 2.2 (95% HPDI 0.5, 5.3) | 4.4 (95% CrI 2.0, 13.4) - 17.3 (3.5, 45.5) | Guinat et al., 2018 |
|  | Pig | Herd | Between herd | - | 1 - 5 | (not reported) | 2-3 | Gulenkin et al., 2011 |
| Armenia 2008 | Boar | Individual | Within pen | 4 | 2 - 9 | (not reported) | 6.1 (95% CI 0.6, 14.5) | Pietschmann et al., 2015 |
|  | Pig, Boar | Individual | Within pen | 4 | 2 - 9 | (not reported) | 5.0 (95% CI 1.4, 10.7) | Pietschmann et al., 2015 |
|  | Pig, Boar | Individual | Between pen | 4 | 2 - 9 | (not reported) | 0.5 (95% CI 0.1, 1.3) | Pietschmann et al., 2015 |
| Genotype IX | Pig | Herd | Between herd | - | 1 month | 1,77 | 1.77 (95% CI 1.74, 1.81) | Barongo et al., 2015 |
|  | Pig | Herd | Between herd | - | 1 month | 0,0059 | 1.58 (range not reported) | Barongo et al., 2015 |
|  | Pig | Herd | Between herd | - | 1 month | 1,90 | 1.90 (95% CI 1.87, 1.94) | Barongo et al., 2015 |
| Not specified | Pig | Herd | Within population | 2.86 - 8.33 days | 1.25 – 100 | 0.001 – 0.3 | 0.8043 – 3.7695 | Shi et al., 2020 |

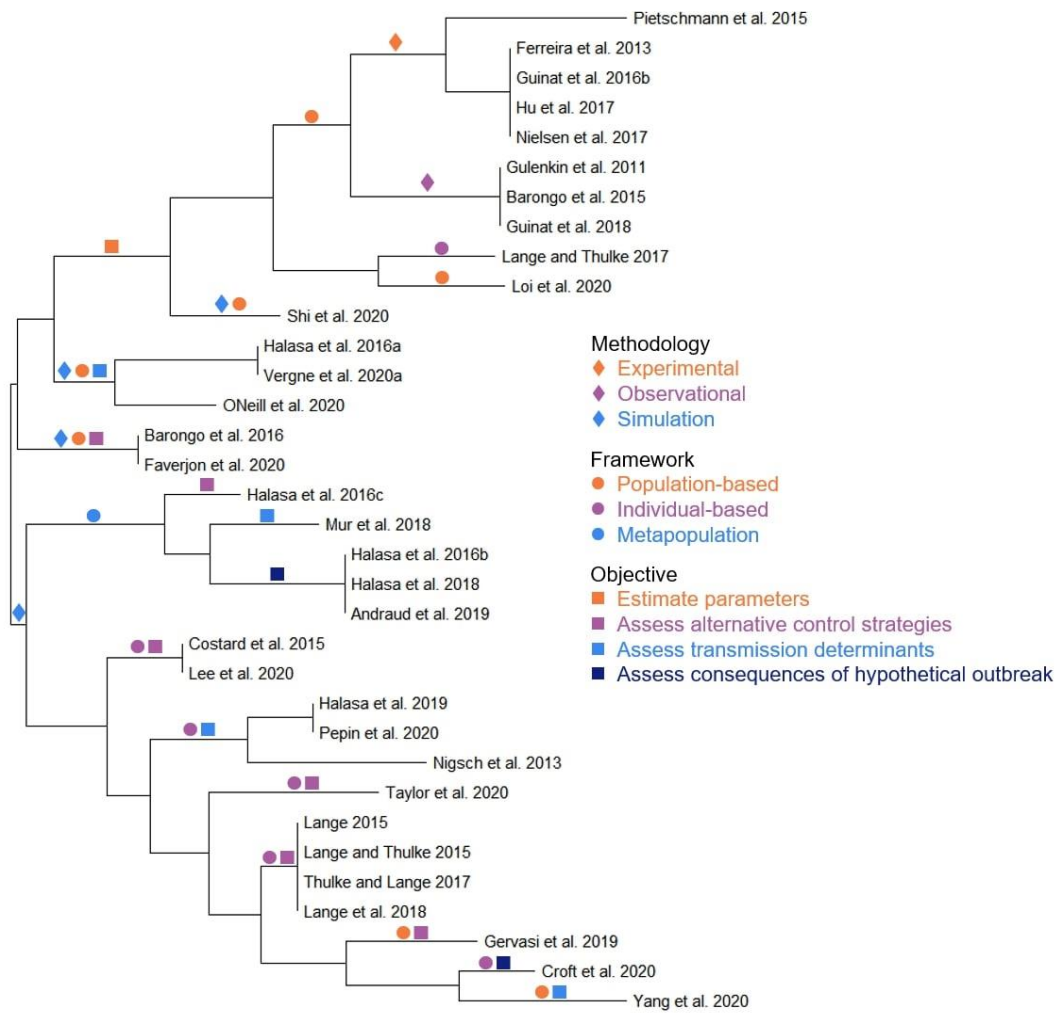

**Figure 3. Distance-based filiation tree of included articles.** Model relationships are based upon species of study (indicated through dashed boxes), and data collection methodology, model framework, and model objective (indicated through color-coded reference symbols overlayed on filiation tree branches). For any symbol that precedes a node (a point of vertical branching), all articles beyond (to the right of) that point contain the characteristic identified by the symbol(s). For example, both *Halasa et al. 2019* and *Lange and Thulke 2017* share an individual-based framework, and tracing further back (to the left) in the tree, use a simulation methodology).

Both the DTU-DADS-ASF and the Be-FAST models relied on simulated live-animal movements and kernel-based distances to model susceptible-infectious contacts between herds. In the DTU-DADS-ASF model, movements (including both animal movements between herds and indirect contacts such as abattoir movements and contact with vehicles and animal health workers) were simulated through series of transmission probabilities parameterized to historical movement frequency data in the represented location (Denmark or France). Distance-based probabilities between herds were used to model local spread. The Be-FAST model also considered direct and indirect contact between herds, using a metapopulation framework to model trade networks and indirect means of spread (Ivorra et al., 2014). Whereas the Be-FAST model used SI infection states within herds, the DTU-DADS-ASF simulation used a modified SEIR model with the infectious state split into sub-clinical and clinical states.

Stochastic, discrete, spatially-explicit individual-based models, mostly focused on assessing alternative control strategies, were the predominant approaches to modelling ASF in wild boar, with the exceptions of Croft et al. (2020) who used a deterministic approach and Gervasi et al. (2019) and Yang et al. (2020) who used deterministic non-spatial population-based models. Of the spatially-explicit individual-based models, unlike in the domestic pig metapopulation models, disease spread was simulated exclusively through movement-based algorithms. For the ASF Wild Boar model (Lange, 2015; Lange and Thulke, 2015; Thulke and Lange, 2017; Lange et al., 2018),

**Table 4. Parameter assumptions**

**Average ASFV infectious period duration**

| Reference | Host | Value | Source |
|---|---|---|---|
| Gulenkin et al., 2011 | Pigs | 1-5 days | FAO, 2009 |
| Barongo et al., 2015 | Pigs | 1 month | Ferreira et al., 2013 |
| Guinat et al., 2016b | Pigs | Min: 3 - 6 days<br>Max: 3 - 14 days | Gabriel et al., 2011; Blome et al., 2012, 2013 |
| Hu et al., 2017 | Pigs | Gamma(mean (days), shape)<br>mean ~ Gamma(10,6.0)<br>shape ~ Gamma(19.3, 2) | Ferreira et al., 2013 |
| Nielsen et al., 2017 | Pigs | 4.5 ± 0.75 days | Guinat et al., 2014 |
| Guinat et al., 2018 | Pigs | Gamma(mean (days), shape)<br>mean ~ Gamma(9.12, 10)<br>shape ~ Gamma(22.20, 5) | Gulenkin et al., 2011; Guinat et al., 2016b; Hu et al., 2017 |
| Lange, 2015; Lange and Thulke, 2015, 2017; Thulke and Lange, 2017; Lange et al., 2018 | Boar | 1 week | Blome et al., 2012 |
| Halasa et al., 2019 | Boar | PERT(1, 5, 7) days | Olesen et al., 2017 |
| Faverjon et al., 2020 | Pig | Uniform (3, 5.5) | Guinat et al., 2016a, 2016b |
| Lee et al., 2020 | Pig | 4-52 weeks | assumed |
| Loi et al., 2020 | Boar | 5-7 days | Gabriel et al., 2011; Blome et al., 2012; Guinat et al., 2016b |
| O'Neill et al., 2020 | Boar | Live boar: 5 days | Gallardo et al., 2015 |
| | | Carcasses: 8 weeks | Carrasco García, 2016; Probst et al., 2017 |
| Pepin et al., 2020 | Pig, Boar | Poisson(5 days) | Blome et al., 2012; Gallardo et al., 2017 |
| Taylor et al., 2020 | Boar | Live boar: PERT(3, 6, 10) days | Gabriel et al., 2011; Guinat et al., 2014 |
| | | Carcasses: PERT(15, 26, 124) days | Morley, 1993; Olesen et al., 2018; Probst et al., 2017; Chenais et al., 2019 |
| Vergne et al., 2020a | Pig | PERT(3, 7, 14) days | Guinat et al., 2016b |
| Yang et al., 2020 | Boar | 5 days | Davies et al., 2017 |

**Beta**

| Reference | Host | Value | Source |
|---|---|---|---|
| Barongo et al., 2016 | Pigs | PERT(0.2, 0.3, 0.5) | Ferreira et al., 2013 |
| Halasa et al., 2016a | Pigs | 0.30 or 0.60 | Guinat et al., 2016b |
| Hu et al., 2017 | Pigs | Gamma(2,2) | Gulenkin et al., 2011 |
| Guinat et al., 2018 | Pigs | Gamma(2, 2) | Gulenkin et al., 2011; Guinat et al., 2016b; Hu et al., 2017 |
| Halasa et al., 2016b, 2016c, 2018 | Pigs | Nuclear, production: PERT(0.14, 0.38, 0.8); Boar, backyard, quarantine, hobby: PERT(0.36, 0.60, 0.93) | Guinat et al., 2016b |
| Mur et al., 2018 | Pigs | Industrial, closed, semi-free: 1.42, Family: 1.85 | Gulenkin et al., 2011 |
| Andraud et al., 2019 | Pigs | Within herd: PERT(0.6, 1, 1.5) | Halasa et al., 2016b |
| Faverjon et al., 2020 | Pig | Within pen: Truncated normal(min, mean, max, sd)(0, 0.6, 14.3, 0.4)<br>Between pen: Truncated normal(0, 0.3, 14.3, 0.2) | Ferreira et al., 2013; Guinat et al., 2016a, 2016b |
| | | Between room: Truncated normal(0, 0.01, 0.1, 0.05) | Assumed |
| Lee et al., 2020 | Pig | Direct contact, indirect contact between small and medium farms: 0.6<br>Indirect contact to large farms: 0.006 | Guinat et al., 2016b |
| Shi et al., 2020 | Pig | 0.001 - 0.3 | Ferreira et al., 2013 |
| Taylor et al., 2020 | Boar | Wild boar to pig: Uniform(0, 0.167)<br>Wild boar to wild boar: PERT(0, 0.167, 0.3)<br>Dead wild boar to wild boar: Uniform(0, 0.167) | Pietschmann et al., 2015 and assumed |
| Vergne et al., 2020a | Pig | PERT(0.2, 0.4, 0.6) | Guinat et al., 2016b |

**Table 4. Parameter assumptions (continued)**

**Average ASFV incubation period duration**

| Reference | Host | Value | Source |
|---|---|---|---|
| Gulenkin et al., 2011 | Pigs | 15 days | OIE, 2008 |
| Nigsch et al., 2013 | Pigs | PERT(3, 5, 13) days | FAO, 2009, Depner personal communication |
| Barongo et al., 2015 | Pigs | 5-15 days | Sanchez-Vizcaino et al., 2015 |
| Costard et al., 2015 | Pigs | Weibull(shape, scale) 2+ (Weibull (1.092, 4.197) (median 5, range 2-19) days | Plowright et al., 1994; Arias and Sanchez-Vizcaino, 2002; Penrith et al., 2004; Sanchez-Vizcaino, 2012 |
| Mur et al., 2018 | Pigs | Poisson(8) | Ferreira et al., 2013; OIE, 2014 |
| Faverjon et al., 2020 | Pig | Gamma(shape, scale) (13.299, 0.3384482) | Ferreira et al., 2012, 2013; Guinat et al., 2016a, 2016b |
| Pepin et al., 2020 | Boar | Poisson(4) days | Blome et al., 2012; Gallardo et al., 2017 |

**Average ASFV latent period duration**

| Reference | Host | Value | Source |
|---|---|---|---|
| Nigsch et al., 2013 | Pigs | 1-2 days | FAO, 2009 |
| Costard et al., 2015 | Pigs | Uniform(1,2) days | Arias and Sanchez-Vizcaino, 2002; Plowright et al., 1994 |
| Pietschmann et al., 2015 | Both | 4 days | Assumed |
| Guinat et al., 2016b | Pigs | 2-5 days | Assumed |
| Barongo et al., 2016 | Pigs | PERT(2.86, 4, 8.3) days | OIE, 2008; FAO, 2008, 2009 |
| Hu et al., 2017 | Pigs | Gamma(mean (days), shape) mean ~ Gamma(4.5, 10) shape ~ Gamma(10, 2) | Ferreira et al., 2013 |
| Nielsen et al., 2017 | Pigs | 3-5 days | Guinat et al., 2014 |
| Guinat et al., 2018 | Pigs | Gamma(mean (days), shape) mean ~ Gamma(6.25, 10) shape ~ Gamma(19.39, 5) | Gulenkin et al., 2011; Guinat et al., 2016b; Hu et al., 2017 |
| Mur et al., 2018 | Pigs | Poisson(2) | Ferreira et al., 2013; OIE, 2014 |
| Halasa et al., 2019 | Boar | PERT(1, 5, 9) days | Olesen et al., 2017 |
| Loi et al., 2020 | Boar | 3.57 days | Gabriel et al., 2011; Blome et al., 2012; Guinat et al., 2016b |
| Shi et al., 2020 | Pig | 2.86 - 8.33 days | Barongo et al., 2016 |
| Vergne et al., 2020a | Pig | PERT(3,4,5) days | Guinat et al., 2016b |
| Yang et al., 2020 | Boar | 4 days | Barongo et al., 2016 |

the model replicated from it (Halasa et al., 2019), and the model by Pepin et al. (2020), this was accomplished using a rasterized spatial habitat grid. In order to avoid raster-associated bias in their model, Croft et al. (2020) elected against a grid-based landscape, instead using a mosaic of irregular polygons scaled to the average wild boar herd range. In all these models, individual animal movements occurred via dispersal and orientation probabilities of each individual animal, followed by upper-bounded number of dispersal steps that could be taken. Unlike domestic pig simulations or the Halasa et al. (2019) and Pepin et al. (2020) wild boar simulations, the ASF Wild Boar individual-based models (Lange, 2015; Lange and Thulke, 2015; Thulke and Lange, 2017; Lange et al., 2018) and Croft et al. (2020) used weekly not daily time steps in their process scheduling.

Three domestic pig models used individual-based frameworks as well, to examine routes of ASF transmission between EU Member States (Nigsch et al., 2013), the efficacy of movement-restriction control measures (Lee et al., 2020), and to assess controlling the silent release of ASF from farms (Costard et al., 2015). For evaluating transmission determinants in the EU, Interspread Plus — a proprietary software program that allows for modelling a variety of animal diseases — used movement-based algorithms to simulate disease spread between herds but did not account for distance-based transmission routes. It was used to model the transmission of ASF both within and between countries. Both pig movements between farms as well as indirect contacts within-country were modelled, followed by simulated export movements. A similar stochastic, discrete, spatially-explicit state-transition model was adapted to the swine network in Vietnam by Lee et al. (2020) — the North American Animal Disease Spread Model (NAADSM). Here, farm-type-dependent contact probabilities and rates simulated animal trade movements. To ascertain the risk of ASF spread secondary to an emergency sell-off of pigs, Costard et al. (2015) developed their own individual-based model. Here, ASF transmission was stochastically simulated within a herd and then coupled to data on the behavior of farmers to determine the risk of ASF spread outside the affected herd.

## (Results) Model insights and assumptions

*Model parameters*

ASF transmission parameters, estimated from models with both individuals and herds acting as the basic epidemiological unit (depending on the study), were often used to parameterize future models — though a variety of other parameter data sources were identified as well (Table 4). This resulted in a range of values being used for ASFV's infectious period, incubation period (the time between infection and clinical signs), and latent period (classically considered as the time between infection and infectiousness, though in Costard et al. (2015) this was defined as infectious without clinical signs) across all models. When ASF data was unavailable, certain parameters had to be adapted from other disease models. Transmission probabilities for pig movements (Nigsch et al., 2013), indirect contacts (Nigsch et al., 2013; Halasa et al., 2016a, 2016b, 2016c; Mur et al., 2018; Halasa et al., 2018; Andraud et al., 2019), and local spread (Halasa et al., 2016a, 2016b, 2016c, 2018; Mur et al., 2018; Andraud et al., 2019) were adapted from CSF studies, as was the range for $R_0$ in Costard et al. (2015). When alternative control strategies were evaluated, some parameters that determined the probability of success of a control measure and the time required for its implementation were adapted from CSF or FMD studies as well (Halasa et al., 2016b, 2016c, 2018; Andraud et al., 2019).

Limited field data for wild boar resulted in the evolution of many assumptions as new information was discovered. Carcass-based transmission was modelled through direct transmission within and between groups first as sex-dependent (Lange and Thulke, 2015), then neither age nor sex-dependent (Lange, 2015; Lange and Thulke, 2017), and then as age-dependent (Lange et al., 2018). Infection probability per carcass was originally parameterized at 20% according to the best-fit model that explained the observed data (Lange and Thulke, 2015). Camera trapping data (Probst et al., 2017) and the results of Lange and Thulke (2017) resulted in this parameter being refined to 2-5% in the subsequent model (Lange et al., 2018). The assumed live infectious periods in the wild boar models were predominantly 5-7 days (Lange, 2015; Lange and Thulke, 2015, 2017; Thulke and Lange, 2017; Lange et al., 2018; Halasa et al., 2019; Loi et al., 2020; O'Neill et al., 2020; Pepin et al., 2020; Taylor et al., 2020), however greater variation was seen among the assumed carcass infectious periods.

In the ASF Wild Boar models, carcass persistence – synonymous with carcass infectivity – was originally statically modelled at 8 weeks (Lange and Thulke, 2015). However, after disease spread was observed and a model was fit, the spread was best explained using a 6-week carcass persistence time (Lange, 2015). Carcass persistence time was further revised to 4 weeks in Lange and Thulke (2017) and Thulke and Lange (2017) (and similarly used in Halasa et al. (2019) in line with field research on vertebrate scavenging behavior from Ray et al. (2014)). The carcass persistence parameter was then further revised to reflect a seasonally-dependent variability in Lange et al. (2018), with persistence times ranging from 4 weeks in the summer to 12 weeks in the winter, in accordance with seasonal differences observed in field research (Ray et al., 2014). This seasonal variability in carcass persistence was also assumed in Pepin et al. (2020). In the later wild boar models, O'Neill et al. (2020) assumed a static carcass infectivity time of 8 weeks, and Taylor et al. (2020) used a PERT distribution of parameters 2, 4, and 18 weeks (specifically: 15, 26, and 124 days), with the latter model also accounting for the probability of carcass removal during the period.

The first wild boar individual-based models (Lange, 2015; Lange and Thulke, 2015) used a 4 km² geographical unit, corresponding to the home range of a wild boar herd, in accordance with ecological data from radio-tracking sessions from Spitz and Janeau (1990) and Leaper et al. (1999). At this unit size there may be some interactions between neighboring herds, though as boar prefer to stay within their home range and interact with their groupmates, long distance movements are consequently mostly related to dispersal of juveniles. The geographical raster was later increased to units of 9 km² (Lange and Thulke, 2017; Thulke and Lange, 2017; Lange et al., 2018) to avoid perfect overlap between the study area and voxel size used in the model (Lange and Thulke, 2017), as necessary for the model objective. The wild boar individual-based model by Halasa et al. (2019), replicated from Lange (2015) and Lange and Thulke (2017), again used 4 km² units. The more recent boar models increase the geographical unit size, with Pepin et al. (2020) using 25 km² grid cells, and Taylor et al. (2020) applying 100 km² cells over the Polish landscape.

Lastly, the timing of viral release varied across the wild boar individual-based models as well. In order to allow population dynamics to become established, virus release was originally set for the first week of the 4th year of simulation run and to 10 hosts in Lange and Thulke (2015). This parameter was adjusted to the beginning of June of the 5th year of simulation (corresponding to the dispersal period for juveniles) and for 25 hosts (Lange, 2015). The next model iterations (Lange and Thulke, 2017; Thulke and Lange, 2017) simulated ASFV release at the end of June of the 4th year of simulation and to 10 hosts, and the following model (Lange et al., 2018) released the infection at the end of June of the 6th year of simulation to 5 hosts. The model described in Halasa et al. (2019) allowed one year for population dynamics to emerge (as evidenced by the dramatic increase in groups in the population graph prior to stabilization), with virus release occurring at the beginning of the second year and to only one random boar. There is no mention of the wild boar population stabilizing before virus introduction. Conversely, Pepin et al. (2020) used a 10-year burn-in period for population dynamics to stabilize prior to ASF release.

*Transmission determinant assessment*

Halasa et al. (2016a) revealed that ASFV's path of transmission through a domestic pig herd is influenced by subclinical animal infectiousness, dead animal residues, and herd size. For spread between pig herds, for the endemic situation in Sardinia where free-roaming unregistered pigs (known as *brado*) complicate eradication efforts, Mur et al. (2018) identified local spread through fomites as the primary transmission route. Brado and wild boar were indicated to play central roles in the occurrence of ASF cases, reinforcing the importance of herd biosecurity in interrupting transmission. On the international scale, it was demonstrated that limited transmission of ASF between EU member nations would occur through swine trade networks prior to disease detection, reinforcing the importance of surveillance measures (Nigsch et al., 2013). Factors influencing the path of transmission of ASFV were also assessed for wild boar in Denmark, where the model showed that the density, size, and location and dispersion of a boar population will affect transmission and circulation of ASF (Halasa et al., 2019). The importance of carcass-based transmission was quantified in Pepin et al. (2020), where it was inferred over half of the transmission events were from infected carcass contact. When observed dynamics of ASF in boar in Europe were modelled – specifically to capture the troughs and peaks of infection and population densities – differences in temperature and scavenger abundance were shown to impact carcass degradation affecting outbreak severity, reinforcing the role of carcasses in epidemic maintenance (O'Neill et al., 2020).

One model explored the role of insect vectors in contributing to disease spread (Vergne et al., 2020a), demonstrating that only a small percentage of ASFV transmission events would be due to stable flies, assuming an average abundance of flies (measured once previously as 3-7 flies per pig). However, as vector abundance increased ten- and twenty-fold, the percentage of transmission due to the insects increased dramatically as well. Transmission was also highly sensitive to blood-meal regurgitation quantity and ASFV infectious dose, indicating areas of necessary further study.

*Alternative control strategy assessment and prediction of consequences of hypothetical outbreaks*

When control strategies were compared and the consequences of outbreaks assessed, Costard et al. (2015) showed that increasing farmers' awareness of and sensitivity of detection to ASF will not reduce the risk of silent release through emergency sales. Barongo et al. (2016) demonstrated that, in a free-range pig population, rapid biosecurity escalation (within 2 weeks of outbreak onset) would significantly decrease the burden of disease. Halasa et al. (2016c) showed that, for industrialized European swine populations, including virological and serological testing of up to five dead animals per herd per week within the perimeter of an outbreak, in addition to established national and EU measures, provided the most effective control strategy. When the consequence of using shorter durations of control zones was assessed, the model predicted such a reduction would greatly reduce economic losses without jeopardizing worsening transmission (Halasa et al., 2018). Conversely, increasing the size of the area under surveillance would offset the increased incurred cost through shortening the epidemic's duration (Halasa et al., 2018). For arresting ASF spread in Vietnam, movement restrictions were used as the control method and it was shown they would have to interdict at least half of all pig movements to be effective. This was problematic as many traders were identified to specifically avoid quarantine checkpoints and sell pigs through illegal means (Lee et al., 2020).

Models that assessed the consequences of hypothetical outbreaks did so for specific industrialized (Danish and French) swine populations and two independent populations of wild boar. The simulations of ASFV spread in the domestic pig compartment only predicted short and small epidemics (mean duration less than one month) in both Denmark and France, with disease spread primarily driven by animal movements and often contained upon implementation of the codified national and EU control strategies (Halasa et al., 2016b; Andraud et al., 2019). As the epidemic could fade out in the inciting herd, some (14.4% of epidemics originating in nucleus herds, 12.1% from sow herds) were predicted to never be detected. Further, the initial outbreak was predicted to have the highest economic cost — more-so than any subsequent outbreaks — due primarily to the ensuing trade restrictions that dwarf the direct costs (Halasa et al., 2016b). In France, due to the pyramidal structure of the swine production system, variation was seen dependent upon the index herd's location in the production pyramid (Andraud et al., 2019). Geographic dispersal of ASF cases was highly dependent on the density of herds where the outbreak initialized, with cases spreading up to 800km from herds in low-density areas. If ASF spread originated from free-range pig herds, as opposed to the top of the production pyramid, it was predicted to potentially affect up to 15 herds. Similar to the results of the assessment of transmission determinants by Mur et al. (2018), local transmission appeared to be the driving route. Among wild boar models, the consequences of concern were the outcome of natural circulation of ASFV in a closed population, where any outbreak was determined to be self-limiting (Croft et al., 2020), and the impact of baiting on disease establishment, where through modelling changes in $R_0$ it was seen that such practice would relatively increase the risk of an ASF epidemic taking hold (Yang et al., 2020).

Wild boar simulations demonstrated the importance of long-term sustained control efforts (i.e. over many generations of wild boar), as the scale of depopulation required for a more rapid solution would likely be untenable (Lange, 2015). As the simulation model parameters were refined with updated evidence, delayed carcass removal (two or more weeks postmortem) was shown to have no effect on curtailing ASF spread; only carcass removal within 1 week (an impractical assumption, given current reported carcass removal rates) was shown to have a positive effect (Thulke and Lange, 2017). This conclusion was expanded in Lange et al. (2018), where successful carcass removal within a core area was shown to reduce the required hunting intensity. A distinction between control methods required for scenarios of focal introduction as opposed to spread from adjacent endemic areas was identified as well: in the case of focal introduction, due to the small size of the affected area, it's possible that a high carcass removal rate could achieve control without the need for intensive hunting (Lange et al., 2018). When surveillance methods were compared, passive surveillance —assuming a 50% carcass detection rate — was shown to be more effective than active surveillance at detecting ASF cases in a small population, however active surveillance was better when both disease prevalence and population density were low (<1.5% prevalence, < 0.1 boar/km²) and the hunting rate was over 60% (Gervasi et al., 2019). When transmission from free-range, outdoor pigs was factored into the spread of ASF from wild boar dispersion, hunting was shown to reduce the number of new cases but not the size of the area at risk, and conversely fencing reduced the size of the region at risk of ASF but not the number of cases (Taylor et al., 2020).

## Discussion

Mechanistic modelling has been a valuable tool for deriving infection parameters, unraveling routes of transmission, assessing alternative control strategies, and determining the consequences of hypothetical outbreaks of ASF. However, despite all that has been elucidated, there is still much research to be done. Existing ASF models are limited in the contexts of their application, their means of evaluating control strategies, and the lack of a bridge between domestic and wild compartments, and attention should be given to resolving these shortcomings.

ASF simulation models, either in domestic pigs or wild boar, have been applied only to a limited number of contexts, despite the epidemic risk faced by all European countries and the insights one could get from mechanistic models to anticipate virus emergence. Simulations of ASF outbreaks in domestic pigs, for the current epidemic of the circulating Georgia 2007/1 isolate, have been published only for two European (Denmark and France) and one Asian (Vietnam) nation. Many differences exist between countries in terms of the type of production system, the distribution of farm types, and the source-nation of imported pigs, preventing the extrapolation of results from one nation to another. Similarly, the presence and distribution of, and control mandates against, wild boar are not uniform between areas, precluding extrapolation of model results outside the area of study. Though the general utility of different control strategies has been indicated, real-world data on wild boar abundance, as difficult as it may be to assess, is needed to facilitate parameterization of these models to real-world scenarios. When the wild boar individual-based models were applied to real-world locations, they were run only at low-population scales: in Denmark where there exists a legal mandate for their elimination, in the Baltic nations but only in the area of the international border, a forest in England, and part of Poland. Of the five-year period in which wild boar models were published, almost half of such publications occurred in the most recent year, 2020. Whereas earlier wild boar models were constructed by only one group, the diversity among the 2020 models is a promising trend in the direction of ASF ecological modelling. However, as the number of individuals being modelled grows the required computing time grows cubically (Keeling and Rohani, 2008), so insightful as these individual-based models may be, presently they may be too computationally expensive to adapt to larger populations in other scenarios or scales.

All models that assessed control strategies did so through comparing a finite set of *a priori* defined interventions. Many control strategies were examined in competition with each other, which is opposed to how they would be actually implemented. For instance, the efficacy of active and passive surveillance for wild boar was considered independently and without the influence of the other in Gervasi et al. (2019), when in reality such methods would be implemented synergistically. While comparing strategies is beneficial for identifying a rank-order efficacy of control methods, this structure does not necessarily determine the most effective combination of all available strategies. Future models should be built to identify the optimal contributions of each control method for achieving specific outcomes (e.g. elimination of ASF cases, or minimizing overall economic impact). This can be achieved by using an objective function where the function inputs are the parameters defining the control strategies (e.g. size and duration of the surveillance and protection zone) and the function output is a measure of the epidemic impact (e.g. total cost of the epidemic) (Rushton et al., 1999; Moore et al., 2010). Optimization algorithms can then be used to examine the space of the input parameter values to find which ones minimize the function output (Hauser and McCarthy, 2009; Moore et al., 2010). It is expected that such modelling output will generate more precise information to policy-makers for designing cost-benefit control strategies.

All models that assess control strategies assume the employed strategies will remain constant over the period of implementation. Due to the evolving nature of epidemics, this is unlikely to reflect real-world conditions. Future models may consider including temporal components to the control strategies, both through parsing by specific pre-defined time points (e.g. optimal control strategies to be used before and after $R_0$ becomes less than 1), as well as via objective functions to identify when is the best time to implement certain strategies (especially with regards to types of surveillance).

Accounting for limitations in the surveillance data used to fit mechanistic models (such as imperfect case detection and delays in reporting) is an important consideration in model development. For instance, many models rely on pig mortality thresholds for detecting ASF, though ASFV could circulate in a herd for almost a month prior to it being detected through such criteria (Guinat et al., 2018). The DTU-DADS-ASF simulation factored in a parameter to account for delays during contact tracing, though detection delays due to imperfect herd-level surveillance (such as from small changes in mortality) was not simulated. Among wild boar, passive carcass detection and under-reporting was a common limitation, as such detection was both seasonally variable and irregular. Taylor et al. (2020) accounted for this through including an "under-reporting factor" in their parameters, while Pepin et al. (2020) fit parameters for this uncertainty using approximate Bayesian computation, though the influence of a lack of negative surveillance data was identified in their analysis. Similarly, when parameters were estimated among wild boar in Sardinia, both non-uniform sampling and a lack of passive surveillance samples were identified as limitations. Though no adjustments were made to address them, the large quantity of data potentially offset the bias, as suggested by the authors. Refining this uncertainty through field studies of wild boar could benefit future models and is worthy of investigation.

Resolving structural uncertainty is another on-going gap in ASF modelling that requires improvement. This uncertainty is demonstrated in multiple ways, such as through the range of values among parameter assumptions and the various routes of transmission (and corresponding scale) that are modelled: where specific routes of indirect transmission may be parameterized in one model another will group all such routes under a single local transmission parameter. Quantifying the contribution of individual indirect routes of transmission to ASF spread is one of many areas for refinement through further research. Whereas uncertainty is a quality inherent to all models, studies have shown that this can be minimized through ensemble modelling, where the results of multiple models are aggregated to generate a common final output. Combinations of models providing the best predictions was demonstrated through the results of the *RAPIDD Ebola forecasting challenge* competition: among a variety of individual- and population-based, stochastic and deterministic, mechanistic and semi-mechanistic models, ensemble predictions routinely performed better than any individual model (Viboud et al., 2018). A similar modelling challenge on ASF was launched in 2020, involving several modelling teams. Though still a work-in-progress, it is anticipated that this exercise will be able to provide similar assessments among ASF models, potentially reinforcing the importance of utilizing synthesized results (INRAE, 2020).

Prior to 2020, there was a noticeable lack of diversity among the existing models. Though the proliferation of models last year helped to offset this imbalance, still over one-third (5/14) of the domestic pig simulations are derived from the DTU-DADS-ASF (and component precursor Halasa et al. (2016a)) model. Similarly, prior to 2020 all but one of the wild boar models were derived from Lange and Thulke's ASF Wild Boar model, and Croft et al. (2020) used epidemiological parameters from Lange and Thulke's model as well. The influx of recent wild boar models by Croft et al. (2020), O'Neill et al. (2020), and Pepin et al. (2020) provided contrasting simulations of wild boar and carcass-based transmission in different outbreak scenarios, helping to diversify the field. This diversity aids in reinforcing the shared conclusions among the different models, such as the importance of combining targeted hunts or culls with active carcass removal to achieve outbreak control while avoiding eradication of the wild boar population (Lange, 2015; O'Neill et al., 2020).

Only one simulation model considered transmission between domestic pigs and wild boar despite differences in the observed transmission pathways between countries. While the individual-based wild boar models not accounting for transmission with domestic pigs may be sufficient for areas with ASF dissemination exclusively in the wildlife compartment, areas where spillover — however intermittently — likely occurs will require models that address this aspect. The one simulation that did consider this inter-compartment transmission relied on contact parameters derived for a free-range savannah-like outdoor farm not typically representative of European swine operations (though the authors accounted for this by assuming such contact as an upper-limit). While this model by Taylor et al. (2020) is a critical step towards a unified ASF model of both domestic pig and wild boar transmission, it also indicates the need to better define the parameters informing wild boar and domestic pig contact risks and rates through further research. Simulation models of hypothetical outbreaks and alternative control strategies that link the domestic and wildlife compartments are critical for informing decision-making. Just as this has been done for multiple other animal diseases such as Aujeszky's disease and hepatitis E (Charrier et al., 2018), foot-and-mouth disease (Ward et al., 2015), and bovine tuberculosis (Brooks-Pollock and Wood, 2015), this should be a priority for all nations at risk of ASF importation.

While mathematical models can provide many insights into disease control, they are far from the only tool available. Recent ASF outbreaks have been successfully controlled without the use of mathematical models, such as in the Czech Republic and Belgium. Multisectoral collaboration between epidemiologists, veterinarians, virologists, ecologists, field-work studies, and expert opinion plays an integral role in ASF control. From model building to outcome validation and decision analysis, experts from these fields should be included to maintain an inclusive multi-faceted approach to ASF modelling.

## 7. Conclusions

With outbreaks across 18 European and 12 Asian nations, ASF has become established as an urgent threat to the global swine industry (ProMED-mail, 2020; Taylor et al., 2020). Mechanistic models have shown much potential for helping to confront this epidemic, however, more modelling studies using empirical data derived from real epidemics are needed, especially for generating better estimates of transmission parameters. As these parameters are integral to designing calibrated intervention plans (such as identifying optimal protection and surveillance zones, or (when available) the fraction of necessary vaccination coverage), and since these parameters have been seen to vary between individual ASF outbreaks, extrapolation of parameters between independent outbreak scenarios is precarious at best. Deriving parameters from Georgia 2007/1 genotype II historical outbreaks beyond the two examinations of the past Russian Federation epidemic (Gulenkin et al., 2011; Guinat et al., 2018) is critical for further refining models to combat the on-going ASF pandemic. Limitations of surveillance systems in obtaining accurate data are an active impediment. Though this is being overcome through more complex modelling and inference techniques (e.g. approximate Bayesian computation), existing labour and workforce limitations hinder field data collection.

Prior to this past year, there was a need to diversify modelling approaches through developing additional frameworks (as almost half of the studies at the time stemmed from one of either two models: DTU-DADS-ASF (Halasa et al., 2016b) and ASF Wild Boar (Lange and Thulke, 2015)), however the large influx of modelling teams in 2020 seeking to address ASF unknowns is a promising direction for the field that will probably be reinforced due to the ASF modelling challenge. In addition, current evidence indicates that spillover events between domestic pigs and wild boar play an important role in ASF outbreaks, and this transmission should be a component of models going forward. Finally, to date, only codified, hypothetical and *a priori* defined interventions were compared. Therefore, moving from intervention comparison to identifying optimized control strategies is critical. Doing so will enable policy-makers to identify the ideal course of action rather than a relatively better option among pre-determined routes.

From a decision point of view, while we promote models to support policy, policy-makers should consider several models together. As ensemble modelling studies have not been performed yet, we recommend using existing models as decision guides only for the specific scenarios modelled. Due to the uncertainty of even basic parameters, and as evidenced in the sensitivity analyses of different models, we do not encourage extrapolating results to non-modelled scenarios (e.g. across national borders). The current modelling body provides excellent insight for addressing ASF transmission at a multitude of scales, and these studies should be referenced as such when forming policy decisions on that level by considering all associated models (i.e. for addressing ASF in Sardinia considering the results of both Mur et al. (2018) and Loi et al. (2019), or when deciding on intra-herd strategy considering the results of Costard et al. (2015), Halasa et al. (2016a), Faverjon et al. (2020), and Vergne et al. (2020a)). For ASF modelers, until uncertain parameters are further refined, we hope our consolidation of parameter assumptions and results will facilitate parameter selection for future models. Addressing all these modelling hurdles is expected to generate more appropriate information, for policy-makers and modellers to contribute to the control of ASF both locally and globally.

## Availability of data and materials

All papers used in this review are listed in the "References" section.

## Competing interests and funding

All authors certify that they have no affiliations with or involvement in any organization or entity with any financial interest or non-financial interest in the subject matter or materials discussed in this manuscript.